\documentclass[structabstract]{aa}  
\usepackage{graphicx}
\usepackage{txfonts}
\usepackage{natbib}
\usepackage{amssymb}
\begin{document}
   \title{Discovery of gamma and X-ray pulsations from the young and energetic PSR~J1357$-$6429 with \emph{Fermi} and \emph{XMM-Newton}}

\author{
M.~Lemoine-Goumard$^{(1,2)}$ \and 
V.~E.~Zavlin$^{(3)}$ \and 
M.-H.~Grondin$^{(4)}$ \and 
R.~Shannon$^{(5)}$ \and 
D.~A.~Smith$^{(1)}$ \and 
M.~Burgay$^{(6)}$
F.~Camilo$^{(7)}$ \and 
J.~Cohen-Tanugi$^{(8)}$ \and 
P.~C.~C.~Freire$^{(9)}$ \and 
J.~E.~Grove$^{(10)}$ \and 
L.~Guillemot$^{(9)}$ \and 
S.~Johnston$^{(5)}$ \and 
M.~Keith$^{(5)}$ \and 
M.~Kramer$^{(9, 11)}$ \and 
R.~N.~Manchester$^{(5)}$ \and 
P.~F.~Michelson$^{(12)}$ \and 
D.~Parent$^{(13)}$ \and 
A.~Possenti$^{(6)}$ \and 
P.~S.~Ray$^{(10)}$ \and 
M.~Renaud$^{(8)}$ \and 
S.~E.~Thorsett$^{(14)}$ \and 
P.~Weltevrede$^{(11)}$ \and 
M.~T.~Wolff$^{(10)}$
}
\authorrunning{LAT collaboration}

\institute{
\inst{1}~Universit\'e Bordeaux 1, CNRS/IN2P3, Centre d'\'Etudes Nucl\'eaires de Bordeaux Gradignan, 33175 Gradignan, France\\
\email{lemoine@cenbg.in2p3.fr}, \email{smith@cenbg.in2p3.fr} \\
\inst{2}~Funded by contract ERC-StG-259391 from the European Community\\ 
\inst{3}~Space Science laboratory, Universities Space Research Association, NASA MSFC VP62, Huntsville, AL 35805, USA\\ 
\email{vyacheslav.zavlin-1@nasa.gov} \\
\inst{4}~Institut f\"ur Astronomie und Astrophysik, Universit\"at T\"ubingen, D 72076 T\"ubingen, Germany\\ 
\email{grondin@astro.uni-tuebingen.de} \\
\inst{5}~CSIRO Astronomy and Space Science, Australia Telescope National Facility, Epping NSW 1710, Australia\\ 
\email{Ryan.Shannon@csiro.au} \\
\inst{6}~INAF - Cagliari Astronomical Observatory, I-09012 Capoterra (CA), Italy\\ 
\inst{7}~Columbia Astrophysics Laboratory, Columbia University, New York, NY 10027, USA\\ 
\inst{8}~Laboratoire Univers et Particules de Montpellier, Universit\'e Montpellier 2, CNRS/IN2P3, Montpellier, France\\ 
\inst{9}~Max-Planck-Institut f\"ur Radioastronomie, Auf dem H\"ugel 69, 53121 Bonn, Germany\\ 
\inst{10}~High Energy Space Environment Branch, Naval Research Laboratory, Washington, DC 20375, USA\\ 
\inst{11}~Jodrell Bank Centre for Astrophysics, School of Physics and Astronomy, The University of Manchester, M13 9PL, UK\\ 
\inst{12}~W. W. Hansen Experimental Physics Laboratory, Kavli Institute for Particle Astrophysics and Cosmology, Department of Physics and SLAC National Accelerator Laboratory, Stanford University, Stanford, CA 94305, USA\\
\inst{13}~Center for Earth Observing and Space Research, College of Science, George Mason University, Fairfax, VA 22030, resident at Naval Research Laboratory, Washington, DC 20375\\ 
\inst{14}~Santa Cruz Institute for Particle Physics, Department of Physics and Department of Astronomy and Astrophysics, University of California at Santa Cruz, Santa Cruz, CA 95064, USA\\ 
}

   \date{July 13th}

 
  \abstract
   {Since the launch of the \emph{Fermi} satellite, the number of known gamma-ray pulsars has increased tenfold. 
    Most gamma-ray detected pulsars are young and energetic, and many are associated with TeV sources. 
    PSR~J1357$-$6429 is a high spin-down power pulsar ($\dot E = 3.1 \times 10^{36}$ erg/s), 
    discovered during the Parkes multibeam survey of the Galactic plane, with significant timing noise typical of very young pulsars. In the very-high-energy domain (E $>$ 100 GeV), H.E.S.S. has reported the detection of the extended source HESS~J1356$-$645 (intrinsic Gaussian width of 12') whose centroid lies 7' from PSR J1357$-$6429.}
   {We search for gamma and X-ray pulsations from this pulsar, characterize the neutron star emission and explore the environment of PSR~J1357$-$6429.}
   {Using a rotational ephemeris obtained with 74 observations made with the Parkes telescope at 1.4~GHz, we phase-fold more than two years of gamma-ray data acquired by the Large Area Telescope aboard \emph{Fermi} as well as those
    collected with \emph{XMM-Newton}, and perform gamma-ray spectral modeling.}
   {Significant gamma and X-ray pulsations are detected from PSR~J1357$-$6429. The light curve in both bands shows one broad peak. Gamma-ray spectral analysis of the pulsed emission suggests that it is well described by a simple power-law of index $1.5 \pm 0.3_{\rm stat} \pm 0.3_{\rm syst}$ with an exponential cut-off at $0.8 \pm 0.3_{\rm stat} \pm 0.3_{\rm syst}$ GeV and an integral photon flux above 100 MeV of ($6.5 \pm 1.6_{\rm stat} \pm 2.3_{\rm syst}$) $\times 10^{-8}$ cm$^{-2}$~s$^{-1}$. 
    The X-ray spectra obtained from the new data provide results consistent with those reported by Zavlin (2007).
    Upper limits on the gamma-ray emission from its potential pulsar wind nebula (PWN) are also reported.}
   {Assuming a distance of 2.4~kpc, the \emph{Fermi} LAT energy flux yields a gamma-ray luminosity for PSR~J1357$-$6429 of $L_\gamma = (2.13 \pm 0.25_{\rm stat} \pm 0.83_{\rm syst})\times 10^{34}$ erg~s$^{-1}$, consistent with an $L_\gamma \propto \sqrt{\dot E}$ relationship. The \emph{Fermi} non-detection of the pulsar wind nebula associated with HESS~J1356$-$645 provides new constraints on the electron population responsible for the extended TeV emission.}

   \titlerunning{Discovery of gamma and X-ray pulsations from PSR~J1357$-$6429 with \emph{Fermi} and \emph{XMM-Newton}}

   \maketitle
%

\section{Introduction}
Since the launch of the \emph{Fermi} Gamma-ray Space Telescope, the number of known gamma-ray pulsars has increased dramatically. The first \emph{Fermi} Large Area Telescope (LAT) catalog of pulsars~\citep{psrcat} also marks 
the birth of new classes of gamma-ray pulsars, such as millisecond~\citep{mspsr} and radio-faint gamma-ray pulsars~\citep{radioquiet}. 
The rest are young pulsars, discovered via an efficient collaboration among gamma-ray, radio and X-ray astronomers. Interestingly, many of the detected 
pulsars are found to be powering pulsar wind nebulae (PWN), and some are associated with TeV sources.

PSR~J1357$-$6429 is a very young, energetic pulsar discovered during the Parkes multibeam survey of the Galactic plane, as 
reported in~\cite{Camilo}. From its dispersion measure of ($127.2 \pm 0.5$) $\rm cm^{-3} \,pc$, the NE2001 model of the Galactic 
electron distribution~\citep{Cordes2002} assigns PSR~J1357$-$6429 a distance of $d= 2.4 \pm 0.6 $~kpc. With a spin period of $166$~ms and
a period derivative of $3.6 \times 10^{-13}$ $\rm{s \ s^{-1}}$, the characteristic age is $\tau = 7300$ yr.

No EGRET source is coincident with PSR~J1357$-$6429.
Recently, Very Large Telescope data were used to search for optical emission
from PSR J1357$-$6429~\citep{mignani}. No counterpart was found, and an upper
limit of V$ \simeq 27$ was determined. Using X-ray \emph{Chandra} data collected in 2005, \cite{zavlin} reported a hint of pulsed emission using an accurate, contemporaneous radio rotation ephemeris. While of low significance (2--3$\sigma$), it is nevertheless more compelling than the non-detection of pulsations 
reported by \cite{esposito}, who folded the same data
with an older ephemeris as a starting point to scan for a rotation period
for which X-ray pulsations might appear. This pulsar's timing noise imposes a large number of period trials,
which in turn diminishes the significance of what was, at best, a weak signal. In this article, we confirm the detection using new \emph{XMM-Newton} data from 2009. 

PSR J1357$-$6429 was highlighted by \cite{psrcat} because of the {\em absence} of
gamma-ray pulsations, and might have become a candidate ``gamma-faint'' pulsar~\citep{romani}. 
Nearby, with a high spin-down power $\dot E = 3.1 \times 10^{36}$ erg~s$^{-1}$, it has $\sqrt{\dot{E}}/d^2$
that is 1\% of Vela's, a threshold above which the LAT detects most pulsars. Here we show that 
the previous non-detection was
due not to beaming or emission physics, but rather because the timing noise typical of
very young pulsars made it difficult to construct an accurate rotation ephemeris after a large glitch early in the \emph{Fermi} epoch. Once enough post-glitch radio observations had been obtained
to build a phase-connected timing model, the gamma-ray pulsations became immediately apparent.
We note that \cite{peliz} reported a weak detection with AGILE. 

Very often, young and energetic pulsars are embedded in compact nebulae powered by relativistic pulsar winds. 
Short X-ray observations were performed in 2005 with \emph{XMM-Newton} and \emph{Chandra} to search for X-ray emission from a PWN~\citep{esposito, zavlin}. Using the \emph{XMM-Newton} data, \cite{esposito} found 
marginal evidence of diffuse emission in the 2--4 keV energy band consisting of a faint elongated structure starting from PSR~J1357$-$6429. 
They also set a 3$\sigma$ 
upper limit on the luminosity of $\sim 3 \times 10^{31}$ $\rm erg \, s^{-1}$ in the 2--10 keV range
on putative diffuse emission 
using the \emph{Chandra} data. 
On the other hand, \cite{zavlin} reported the detection of tail-like emission associated with the pulsar with a 0.5--10 keV luminosity of 
$\sim 2.5 \times 10^{31} \rm erg \, s^{-1}$. In the very high energy domain (E $>$ 100 GeV), the H.E.S.S. data show significantly
extended emission (intrinsic Gaussian width of $0.20^{\circ} \pm 0.02^{\circ}$) lying close to PSR~J1357$-$6429 \citep{renaud}. 
In this article, a GeV upper limit is derived on the emission from the possible PWN observed in the TeV band, providing new constraints on the leptonic population responsible for the very high energy emission.\\


\section{LAT description and observations}
\label{lat}
The LAT is a gamma-ray telescope that detects photons by conversion
into electron-positron pairs in the energy range between 20 MeV to more than 300 GeV, as described by~\cite{Atwood et al. 2009}.
It is made of a high-resolution converter/tracker (for direction measurement of the incident gamma rays), a CsI(Tl) crystal calorimeter (for energy measurement) and 
an anti-coincidence detector to identify the background of charged particles. 
Compared to EGRET, the LAT has a larger effective area ($\sim$ 8000 cm$^{2}$ on-axis above 1~GeV), 
a broader field of view ($\sim$ 2.4 sr) and superior angular resolution ($\sim$ 0.6$^{\circ}$ 68$\%$ containment at 1 GeV for events converting in the front section of the tracker). 
The on-orbit calibration is described in \cite{OnorbitCalib}.

The analysis used 29 months of data collected starting August 4, 2008, and extending until January 15, 2011. 
We selected events with energies greater than 0.1 GeV but
excluded those with zenith angles larger than 100$^{\circ}$, to minimize contamination from secondary gamma-rays from the Earth's atmosphere~\citep{FermiAlbedo}. 
We selected events in the {\emph{Source}} class, as described in~\cite{Atwood et al. 2009}. {\emph{Source}} events optimize the trade-off between gamma-ray detection
efficiency and residual charged-particle contamination in a manner well-suited to localized, persistent sources, as distinct from the {\emph{Transient}} or {\emph{Diffuse}} classes.
We used the most recent analysis version, called ``Pass~7'', 
documented at the \emph{Fermi} Science Support Center~\footnote{FSSC: http://fermi.gsfc.nasa.gov/ssc/data/}. 
%

\section{Radio timing observations}
\label{radio}
As part of the space- and ground-based pulsar timing campaign supporting \emph{Fermi} \citep{Smith08}, the pulsar 
PSR~J1357$-$6429 is being observed at the 64-m Parkes radio telescope \citep{wel10}.
A total of 74 observations spanning MJDs 54220 to 55575 were made in a band centred at $1369$~MHz. In this band, the pulsar has a flux density of $440~\mu$Jy~\citep{Camilo}. Observations were made using the central beam of the 20~cm Multibeam Receiver~\citep{staveley} and recorded using one of the digital filterbanks (DFBs) built by the Australia Telescope National Facility. The DFBs employed real time pulse phase folding to average data into $30$~second sub-integrations using $256$~MHz of total bandwidth. After calibration, pulse profiles for $\sim 5$~minute observations were cross correlated with a template to obtain times-of-arrival. The TEMPO2 timing package \citep{hob06} was 
then used to build the timing solution, taking into account a significant glitch at MJD 54786 and strong timing noise prevalent in young pulsars. The glitch was modeled by including jumps in spin
phase, spin frequency, and spin frequency derivative at the glitch epoch. Higher order spin frequency derivatives and harmonically related sinusoids (Hobbs et al 2004) were used to model the
timing noise. The post-fit rms is 1.4~ms, or 0.8~\% of a pulsar rotation. 
This timing solution will be made available through the FSSC$^1$.

\section{X-ray observations and data reduction}
\label{Xray}
\emph{XMM-Newton} observed PSR~J1357$-$6429 on October 14--15, 2009 with the EPIC-pn instrument operated in Small Window mode and two EPIC-MOS detectors in Full Window mode (for 55.5 and 78.1 ks 
effective exposures, respectively). The data were reprocessed and reduced with the \emph{XMM-Newton} Science Analysis Software (SAS v.~11.0).

The pulsar was also observed on October 8--9, 2009 with the \emph{Chandra} ACIS-I instrument operated in Timed Exposure (Very Faint) mode for a 59.2 ks effective exposure. The \emph{Chandra} Interactive Analysis of
Observations software (CIAO v.~4.3; the calibration database CALDB v.~4.4.3) was used to generate the ``level 2'' event file for this observation.

\section{Analysis and results}
\subsection{Gamma- and X-ray Light Curves}
\label{pulsar}
PSR~J1357$-$6429 is located $2.5^{\circ}$ below the Galactic plane where the diffuse gamma radiation is still intense. 
For the timing analysis, we took photons within a radius of $0.8^{\circ}$ with respect to the radio pulsar position ($l=309.922705^{\circ}$, $b=-2.514079^{\circ}$) using an energy-dependent cone of radius $\theta_{68} \leqslant$ $ \rm \sqrt{(5.3^{\circ} \times (\frac{E}{100~MeV})^{-0.745})^2 + 0.09^2}$. This choice takes into account the instrument performance and maximizes 
the signal-to-noise ratio over a broad energy range~\citep{Atwood et al. 2009}. The arrival times of events were 
corrected to the Solar System Barycenter using the JPL DE405 Solar System ephemeris \citep{standish98}, 
and the events were folded using the radio ephemeris from Parkes, using the TEMPO2 {\em fermi} plugin.

\begin{figure}
\begin{center}
\includegraphics[width=9cm]{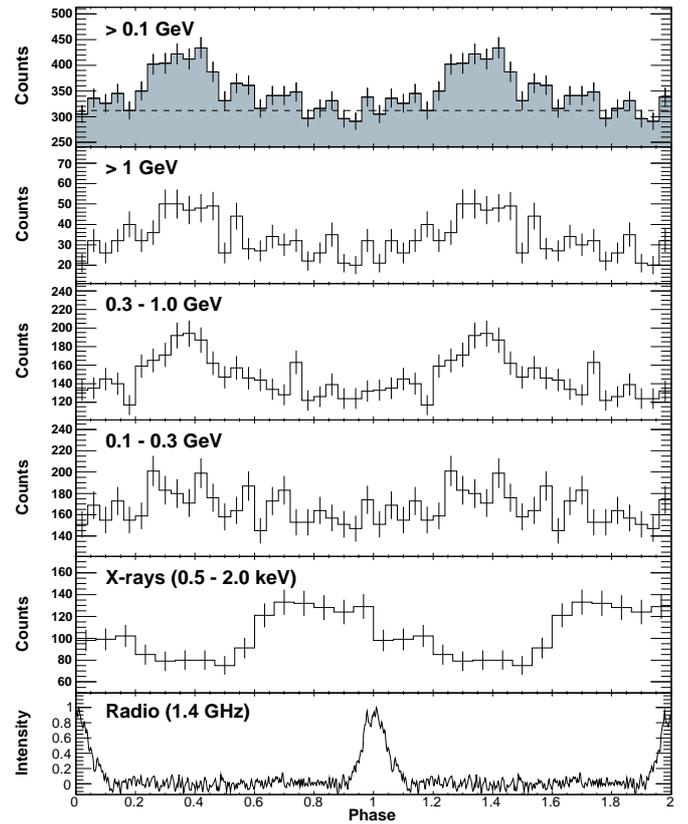}
\caption{
\textbf{Top panel:}~Phase-aligned histogram of PSR~J1357$-$6429 above 0.1 GeV and within an energy-dependent circular region as defined in section~\ref{pulsar}. Two rotations are plotted with 25 bins per period. The dashed line shows the background level estimated using two nearby circular regions as described in Section~\ref{pulsar}. 
\textbf{Three following panels:}~Phase histograms for PSR~J1357$-$6429 in the three indicated energy ranges, each with 25 bins per pulse period. \textbf{Second panel from bottom:}~X-ray pulse profile extracted from the \emph{XMM-Newton} data of 2009 in the 0.5 -- 2 keV energy band. Two rotations are plotted with 15 bins per period.
\textbf{Bottom panel:}~Radio pulse profile based on Parkes observations at a center frequency of 1.4\,GHz with 256 phase bins~\citep{wel10}. \label{fig1} }
\end{center}
\end{figure}

A total of 8707 gamma rays remain above 100 MeV in our energy-dependent circular region, and among them $915 \pm 161$ pulsed photons after background subtraction. We estimated the background level represented by the dashed horizontal line in Figure~\ref{fig1} (top panel) using two circular regions 1.5$^{\circ}$ from PSR~J1357$-$6429 and at the same Galactic latitude. Nearby sources were avoided, and we normalized to the solid angle of the source region. 
An H-test value of 89.6 is obtained above 100 MeV, corresponding to a pulsed detection significance of $\sim$8 $\sigma$ \citep{Htest}. Figure~\ref{fig1} shows the 25 bin gamma-ray light curve (top panel) as well 
as the 1.4~GHz radio profile (bottom panel) derived from observations with the Parkes radio telescope~\citep{wel10}. The gamma-ray peak is offset from the radio pulse by $0.37 \pm 0.03 \pm 0.01$ 
according to a Lorentzian fit with a full width at half maximum (FWHM) of $\sim 0.25$. The first phase uncertainty arises from the gamma-ray fit. The uncertainty in DM of 0.5 pc~cm$^{-3}$  causes an uncertainty in the extrapolation of the radio pulse time-of-arrival to infinite frequency of 0.01 in phase. There is no 
significant evolution of the gamma-ray peak with energy as can be seen from Figure~\ref{fig1} (middle panels). We defined the `off-pulse interval' as the pulse minimum between 0.8 and 1.1 in phase.

Figure~\ref{fig1} also shows the pulse profile obtained from the {\em XMM-Newton} EPIC-pn data with 6~ms time resolution. For the timing analysis, 1543 events were extracted 
(with 270 background counts) from a $15''$-radius circle centered at the pulsar position in the 0.5-2 keV range to maximize the signal-to-noise ratio. The 
{\em photons} plugin\footnote{{\tt
http://www.physics.mcgill.ca/$\sim$aarchiba/ \\ photons\_plug.html}}
TEMPO2 package was used to assign a phase to each selected photon. This allowed us
 to use the full complexity of the ephemeris parameters including 
 those for the glitch mentioned in Section~3 in the timing model.
Performing an unbinned maximum likelihood fit of the phases to a constant plus single Gaussian we obtained an intrinsic pulsed fraction of $0.56 \pm 0.13$
(accounting for the contribution of the diffuse emission 
surrounding the pulsar; see Section~5.2),
with a Gaussian of width $0.48 \pm 0.18$ centered at phase $0.83 \pm 0.03$.  
The H-test yields a value of $H = 52.42$  corresponding to the probability of a chance occurrence in 
a single trial $p=1.6 \times 10^{-11}$, or a 6.7$\sigma$ detection significance. 
The main contribution to the $H$ value comes from 
the fundamental and first harmonics, in accordance with the shape of the X-ray light curve with a single broad pulse. Such a pulse profile can be produced
by a thermal radiation emitted from the neutron star surface (see Section~5.2).
This result unambiguously confirms the previous evidence of pulsations found by~\cite{zavlin} in the {\em Chandra} data taken in 2005.

\subsection{Spatial and spectral analysis of the X-ray data}
\label{xana}
\begin{figure}
\begin{center}
\includegraphics[width=9cm]{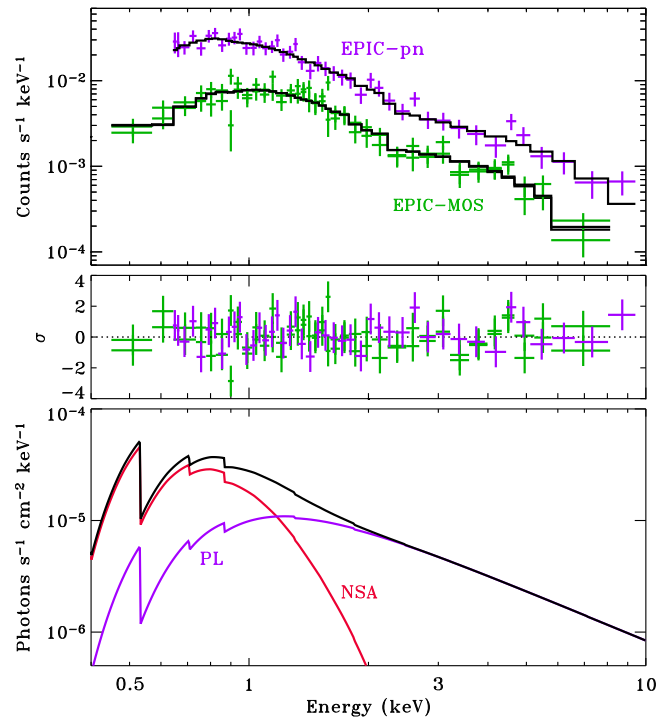}
\caption{Spectra of PSR J1357$-$6429 detected with the \emph{XMM-Newton} EPIC instruments (upper panel) fitted with a two-component, NSA-plus-PL, model
(lower panel). The middle panel shows residuals in the spectral fit.
\label{figspec}
 }
\end{center}
\end{figure}

We used the \emph{XMM-Newton} data from 2009
 to derive the spectrum of PSR~J1357$-$6429. Photons were extracted from a $20''$-radius circle centered at the pulsar position in the EPIC-pn and MOS data. Background was evaluted from similar regions in the vicinity of the pulsar. The estimated numbers of source counts are $2266\pm 72$, 
 $1000\pm 40$ and $996\pm 39$ for the EPIC-pn, MOS1 and MOS2 data sets 
 (respectively).
 The instrumental responses were generated with the SAS {\em rmfgen} and 
{\em arfgen} tools. The purely non-thermal interpretation of the pulsar's X-ray
emission
can be most likely ruled out since the single power-law (PL) fit to the
obtained spectra results in a best estimate of the hydrogen column density,
$n_{\rm H} < 2\times10^{19}$ cm$^{-2}$, significantly lower than the value 
suggested 
by the pulsar dispersion measure, $n_{\rm H} \simeq 4\times10^{21}$ cm$^{-2}$
(see \cite{zavlin} for details). Adding a thermal component, blackbody (BB) or
neutron star magnetized atmosphere model (NSA; see \cite{zavlin09} for a 
review) yields reasonable values of $n_{\rm H}$ and
provides good fit quality (with $\chi^2_\nu=0.9$--1; 
see Figure~\ref{figspec}).  
Generally, the results derived with the two-component model fits to 
the new data are consistent with those reported in \cite{zavlin}.
For example, the NSA-plus-PL fit gives the estimate on the neutron star effective temperature $T_{\rm eff}=(0.95 \pm 0.05)\times 10^6$ K
(assuming the standard star mass $M=1.4\,M_\odot$
and radius $R=10$ km, and the distance $d=2.4$ kpc). 
The non-thermal component is well described by the PL model with a photon index $\Gamma = 1.43 \pm 0.14$ and (unabsorbed) flux 
$F_X\simeq 2.0\times 10^{-13}$ ergs cm$^{-2}$ s$^{-1}$ in 
0.5--10 keV (this range is used for the estimates on all non-thermal fluxes
below), 
whereas the hydrogen column density is $n_{\rm H}$ =  $(3.9 \pm 0.4) \times 10^{21}$~cm$^{-2}$. 

\begin{figure}
\begin{center}
\includegraphics[width=6.5cm]{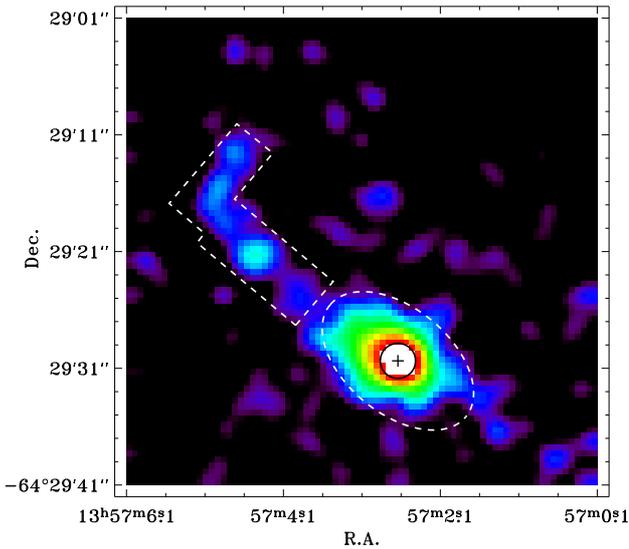}
\vspace*{0.7cm}
\caption{\emph{Chandra} ACIS-I image of 
PSR J1357$-$6429 (in 0.5-10 keV) surrounded by an extended structure. 
The dashed contour indicates the jet-like feature
elongated to the North-East direction.
The cross and the $1\farcs5$-radius circle give the pulsar radio position.
The elliptical contour (with $7''$ and $4''$ major and minor
semi-axes) indicates the diffuse emission surrounding the
pulsar (minus that located within the circle).
\label{figchandra}
 }
\end{center}
\end{figure}

In addition to the results yielded by the new {\em XMM-Newton} observation,
the \emph{Chandra} ACIS-I data of 2009
reveal diffuse emission both surrounding PSR~J1357$-$6429 and extending $15''$--$20''$ 
from the pulsar in the north-east direction
(see Figure~\ref{figchandra}). Despite the small number of counts ($55 \pm 9$) estimated in the elongated and bent feature (outlined with 
the dashed contour in Figure~\ref{figchandra}), its detection is significant
at a 6.1$\sigma$ level. 
The spectrum of this feature, extracted with the CIAO tool {\em specextract}
that yields both the spectra and instrumental responses,
is fitted with a PL model of $\Gamma=1.51\pm 0.28$
and unabsorbed flux $F_X\simeq 0.2\times 10^{-13}$ ergs cm$^{-2}$ s$^{-1}$. 
This confirms the earlier detection of a potential PWN reported 
by~\cite{zavlin}. 
The diffuse emission surrounding the pulsar (denoted with the elliptical
contour in Figure~\ref{figchandra}, with the estimated number of
$224\pm 15$ source counts),
is characterized by a PL spectrum of $\Gamma=1.48\pm 0.18$ and 
$F_X\simeq 0.7\times 10^{-13}$ ergs cm$^{-2}$ s$^{-1}$.
A detailed analysis of the detected nebula is outside the scope of this 
paper but see \citet{Chang2011} for more results on the X-ray PWN.

\begin{figure}
\begin{center}
\includegraphics[height=6.1cm]{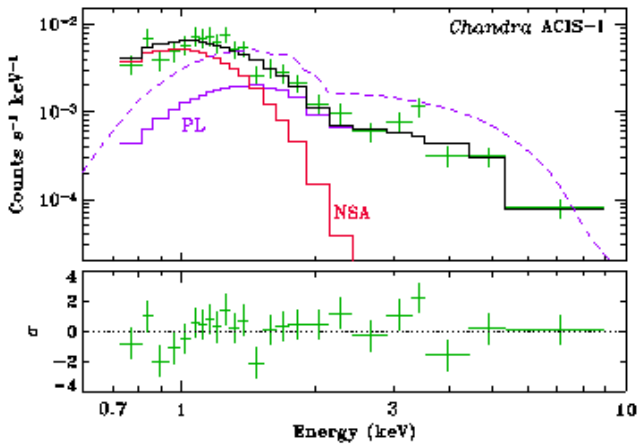}
\caption{Spectrum of PSR J1357$-$6429 detected with the \emph{Chandra} ACIS-I instrument (crosses in the upper panel) and fitted with a NSA-plus-PL model
(solid curves).
The thermal (NSA) component, as well as the hydrogen column density, is
the same as in Figure~2. The dashed curve indicates the PL component derived
from the EPIC spectra.
 The lower panel shows residuals in the spectral fit.
\label{figchandra2}
 }
\end{center}
\end{figure}

But it should be noted
 the diffuse emission
 is included in the extraction region used to derive the X-ray spectra 
 from the \emph{XMM-Newton} EPIC instruments. Therefore, a fraction of the
 non-thermal flux measured from the EPIC data
 is contributed by the 
 PWN.
This is clearly seen in Figure~\ref{figchandra2},
showing the pulsar's spectrum extracted from the $1\farcs5$-radius circle centered at its 
radio position in the \emph{Chandra} ACIS-I data 
($456\pm 21$ source counts; see Figure~\ref{figchandra}). While the thermal component 
(and the hydrogen column density) is well described by the NSA model yielded by the 
\emph{XMM-Newton} data, the non-thermal component is fitted with a PL model of 
photon index $\Gamma=1.45\pm 0.16$ and unabsorbed X-ray flux 
$F_X\simeq 0.9\times 10^{-13}$ ergs cm$^{-2}$ s$^{-1}$
(with $\chi^2_\nu=1.1$; Figure~\ref{figchandra2}). 
We note that the sum of the non-thermal fluxes
estimated in the \emph{Chandra} data
for the pulsar, the surrounding emission and the elongated tail
match well that derived from the \emph{XMM-Newton} spectra.

Another interesting point to mention is that the elongation of
the detected X-ray PWN, about $60^\circ$ to North-East, may indicate
the direction of the pulsar's proper motion \citep[see also][]{mignani}. 

\subsection{\emph{Fermi} LAT spectra and phase-averaged flux}
\label{spectrum}
The phase-averaged spectrum of PSR~J1357$-$6429 was obtained using \emph{gtlike}, a maximum likelihood spectral analysis~\citep{mat96} 
implemented in the \emph{Fermi} Science Tools$^1$. This tool fits a source model to the data. 
The model includes the object under study as well as other gamma-ray sources
in a region extending well beyond the LAT's angular resolution. 
Sources near PSR~J1357$-$6429 exceeding the background with statistical significance larger than $5 \, \sigma$ are extracted from the \emph{Fermi} LAT 
2FGL catalog \citep{SecondCat}. Galactic diffuse emission is modeled using the ring-hybrid model {\it ring\_2year\_P76\_v0.fits}. 
The instrumental background and the extragalactic radiation are described by a single isotropic component with the spectral shape in
the tabulated model {\it isotrop\_2year\_P76\_source\_v0.txt}. The likelihood tool \emph{gtlike} further requires tables of the Instrument Response Functions (IRFs),
which are the energy and direction dependent angular resolution, effective area, and energy resolution. We used the P7\_V6 IRFs.
These new models and their detailed description will be available through the FSSC.

\begin{figure}
\begin{center}
\includegraphics[width=9cm]{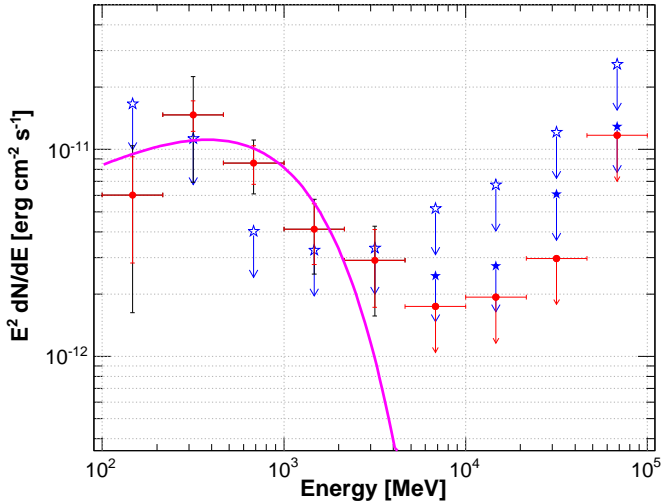}
\caption{
Spectral energy distribution of PSR J1357$-$6449 in gamma rays.
  The LAT spectral points (in red dots) are obtained using the maximum likelihood
  method \emph{gtlike} described in Section~\ref{spectrum} in 9
  logarithmically-spaced energy bins between 100 MeV and 100 GeV. The statistical errors are shown
  in red, while the black lines take into account both the statistical
  and systematic errors as discussed in Section~\ref{spectrum}. The magenta solid line presents the result obtained by
  fitting an exponentially cut off power law to the data in the 100 MeV~--~100~GeV energy range
  using a maximum likelihood fit. A 95~\% C.L. upper limit is computed
  when the statistical significance is lower than 3~$\sigma$ (red dots). Blue filled (open) stars represent the 95~\% C.L. upper limits derived assuming a Gaussian of $0.2^{\circ}$ at the position of HESS~J1356-645 in the whole signal (in the off-pulse). \label{fig2} }
\end{center}
\end{figure}

Assuming a power-law with an exponential cut-off spectral shape, the best fit result is obtained for a spectral index of 
1.5 $\pm$ 0.3 $\pm$ 0.3 with a cut-off at $0.8 \pm 0.3 \pm 0.3$~GeV and an integral photon flux above 100~MeV of (6.5 $\pm$ 1.6 $\pm$ 2.3) $\times 10^{-8}$ cm$^{-2}$~s$^{-1}$. The integral energy flux is (1.9 $\pm$ 0.2 $\pm$ 0.8) $\times 10^{-5}$~MeV~cm$^{-2}$~s$^{-1}$. The first error is statistical,
while the second represents our estimate of systematic effects as discussed below. To check the assumption of a cut-off energy in the spectrum, we have also fit the same dataset with a simple power-law. The spectral model using an exponential cut-off is better constrained with a difference between the log likelihoods of $\sim 4\,\sigma$, disfavoring the power-law hypothesis. Figure~\ref{fig2} shows the spectral energy distribution of PSR~J1357$-$6429 as seen by \emph{Fermi}. The \emph{Fermi} LAT spectral points were obtained by dividing the 100 MeV~--~100 GeV range into 9 logarithmically-spaced energy bins and performing a maximum likelihood spectral analysis in each interval,
assuming a power-law shape for the source. These points
are overlaid with the exponentially cut off power law fit over the total energy range.

Two main systematic uncertainties can affect the LAT flux estimation for a point source: uncertainties on the Galactic diffuse background and on the effective area. 
The dominant uncertainty at low energy comes from the Galactic diffuse emission since PSR~J1357$-$6429 is located only $2.5^{\circ}$ from the Galactic plane. 
By changing the normalization of the Galactic diffuse model artificially by $\pm 6$\% as done in \cite{w49}, we estimate the systematic error on the integrated flux of the 
pulsar to be 50\% below 500 MeV, 20\% between 500 MeV and 1 GeV, and 7\% above 1~GeV. 
The second systematic is estimated by using modified IRFs whose effective areas bracket those of our nominal IRF. 
These `biased' IRFs are defined by envelopes above and below the nominal energy dependence of the effective area by linearly 
connecting differences of (10\%, 5\%, 20\%) at log(E) of (2, 2.75, 4), respectively.  
We combine the errors in quadrature to obtain our best estimate of the total systematic uncertainty at each energy, and propagate through to the fit model parameters.

As can be seen from Figure~\ref{fig2}, no significant emission is detected above 4.5 GeV at the position of PSR~J1357$-$6429. In very high energy gamma rays, the H.E.S.S. experiment has detected significant emission close to the pulsar and significantly extended with an intrinsic 
Gaussian width of $0.20^{\circ} \pm 0.02^{\circ}$. As a first search for unpulsed emission from the TeV source HESS~J1356$-$645, we fitted a Gaussian 
of $0.2^{\circ}$ to the data in the 
energy band 4.5\ --\ 100 GeV where no pulsed emission is detected. No significant signal could be observed and we derived 95\% Confidence Level (CL) upper 
limits on the flux in the 4 logarithmically-spaced energy bins between 4.5 and 100 GeV assuming a power-law shape for the source with fixed spectral index of $2$. In a second step, we used the off-pulse data selecting photons 
in the 0.8 $-$ 1.1 phase interval and fit the same Gaussian of $0.2^{\circ}$ to data in the whole energy band 100 MeV \ --\ 100 GeV. Again, 
no significant emission could be detected and we derived 95\% CL upper limits on the flux in the 9 logarithmically-spaced energy bins (assuming the same spectral index). These upper limits are presented in Figure~\ref{fig2}.

\section{Discussion}
\cite{Johnston2006} measured the polarization profiles for 14 young pulsars at Parkes. 
For PSR J1357$-$6429 the polarization position angle $\Psi$ increases with a constant slope of 
${d\Psi \over d\phi} = 2.7$ across the radio pulse. In the simplest RVM scenario \citep[Rotating 
Vector Model,][]{RVM}, $ \sin \alpha = {d\Psi \over d\phi}\sin \beta $,
hence the angle $\beta$ between the magnetic axis and the line-of-sight is small,
$\beta < 22^{\circ}$. The lack of an inflection in $\Psi$ vs $\phi$ and the narrow $\phi$ range of the pulse
make a full RVM fit unconstraining. 

PSR J1357$-$6429's gamma-ray profile resembles that of PSR J2229+6114~\citep{psrj2229}, PSR J0248+5832~\citep{psrj0248}, and
PSR J1718$-$3825~\citep{psrj1718} in having a single, wide peak offset from the radio pulse by a bit less than a half-rotation. 
The ``Atlas'' update by \citet{AtlasII} provides gamma-ray profiles calculated for different emission models, over a
coarse grid of inclination angles $\alpha$ and $\zeta = \alpha + \beta$, and for 4 gap widths $w$ approximated by
the efficiency, $\eta = L_\gamma/\dot E$.
We find $L_\gamma = (2.1 \pm 0.3 \pm 0.8)\times 10^{34}f_\Omega (d / 2.4 \rm \, kpc)^2$ erg~s$^{-1}$ and $\eta = 0.01 f_\Omega (d / 2.4 \rm \, kpc)^2$ above 100~MeV,
with the beaming correction factor $f_\Omega$ as defined by \cite{AtlasII}.
Two profiles therein resemble the data. They are for $w = 0.01$, $\alpha = 55^\circ$, 
and $\zeta = 35^\circ$ and $50^\circ$ (adjacent plots in the coarse $(\alpha, \zeta)$ grid).
This is consistent with small $\beta = \zeta - \alpha$ suggested by the RVM.
For these geometries, the Atlas predicts $f_\Omega \simeq 1$, that is, a small correction to $L_\gamma$.
This luminosity is within the distance-dominated scatter of the $L_\gamma \propto \sqrt{\dot E/10^{33}}$ rule 
shown by \cite{psrcat}, with $\dot E$ in units of erg s$^{-1}$.

The new X-ray data collected from PSR~J1357$-$6429 with {\em XMM-Newton}
and {\em Chandra} in 2009 strongly suggest that the bulk of the pulsar flux at soft X-ray energies
(below about 1.5 keV) is of thermal origin, making this pulsar the second
youngest neutron star with a detected thermal component known at this time (see
\cite{zavlin} for details). As the X-ray pulsed flux shown in Figure~1
is extracted in the 0.5-2 keV range, it mostly comes from 
thermal radiation. It explains the sine-like shape of the light curve
with the single broad pulse that is characteristic of the majority of X-ray
pulse profiles observed from pulsars with dominating thermal emission.
It is also consistent with the phase shift between the pulses
in the gamma-ray and X-ray fluxes --- the pulses originate at different sites. 
The thermal component is emitted from the neutron star surface, whereas
the high-energy radiation is generated far from the surface
(e.~g., in the pulsar magnetosphere). Yet the origin of the thermal
component remains unclear. It may be emitted from 
 a small hot area (polar caps) of a radius of 1--2 km on the pulsar's surface,

 as suggested by the interpretation involving the BB model,
or from the entire surface as indicated by the fits with the NSA models.
To discriminate between these two interpretations, one should model the
pulsed flux, as has been done for X-rays detected from a number of
millisecond pulsars \citep[e.g., ][]{zp98, bogd08}. This modeling should
take into account the effects of the neutron star geometry (orientation
of the pulsar spin and magnetic axes with respect the line of view),
surface magnetic and temperature distributions, intrinsic properties (anisotropy) of the thermal emission, and the gravitational bending of
photon trajectories near the star surface. If future X-ray observations 
provide much better photon statistics, such an analysis can be supported
by phase-resolved spectroscopy and energy-resolved pulse modeling. 
At present, we can only speculate that the detected thermal flux is
intrinsically anisotropic, as predicted by the atmosphere models,
otherwise the effect of the gravitational bending 
 would strongly suppress the pulsations.
 

PSR~J1357$-$6429 also lies within the extended TeV source HESS~J1356$-$645, proposed as its associated PWN~\citep{renaud}. No significant signal is detected using \emph{Fermi} LAT data above 4.5 GeV, nor in the off-pulse interval below 4.5 GeV. In the context of the leptonic model proposed by \cite{renaud}, our upper limits constrain the spectrum of the electron population responsible for the extended TeV emission. The best model reproducing both the radio, X-ray and TeV data as well as the non-detection by \emph{Fermi} is obtained for a magnetic field of 4 $\mu$G (similar to what is seen in other evolved TeV PWNe) and an electron spectrum characterized with a spectral index of 2.4, a minimum energy of 10~GeV, a cut-off energy of 100~TeV and a total energy of $4 \times 10^{47}$~erg~\citep{renaud}. This energy is only $\sim$10\% of the total kinetic energy since birth of PSR~J1357$-$6429, assuming an initial period
  of ~80 ms.

\begin{acknowledgements}
The \emph{Fermi} LAT Collaboration acknowledges generous ongoing
support from a number of agencies and institutes that have
supported both the development and the operation of the LAT as well as 
scientific data analysis. These include the
National Aeronautics and Space Administration and the Department
of Energy in the United States, the Commissariat \`a
l'Energie Atomique and the Centre National de la Recherche
Scientifique / Institut National de Physique Nucl\'eaire et de
Physique des Particules in France, the Agenzia Spaziale Italiana,
the Istituto Nazionale di Fisica Nucleare, and the Istituto
Nazionale di Astrofisica in Italy, the Ministry of Education,
Culture, Sports, Science and Technology (MEXT), High
Energy Accelerator Research Organization (KEK) and Japan
Aerospace Exploration Agency (JAXA) in Japan, and the K.
A. Wallenberg Foundation and the Swedish National Space
Board in Sweden.
Additional support for science analysis during the operations 
phase from the following agencies is also gratefully acknowledged: 
the Instituto Nazionale di Astrofisica in Italy and the Centre National d'\'Etudes Spatiales in France.\\
The Parkes radio telescope is part of the Australia Telescope which is funded by the Commonwealth Government for operation as a National Facility managed by CSIRO. We thank our colleagues for their assistance with the radio timing observations.\\
VEZ is grateful to Nataliya Ivanova for discussions and support.
\end{acknowledgements}

\end{document}